\documentclass[twocolumn,aps,prb,showpacs,amsmath,superscriptaddress,longbibliography,notitlepage]{revtex4-1}
\usepackage{amssymb}
\usepackage{mathrsfs}
\usepackage{graphicx}
\usepackage{float}
\usepackage[caption=false]{subfig}
\usepackage[pdftex,colorlinks=red]{hyperref}
\usepackage{verbatim}
\usepackage{txfonts}
\usepackage{amsmath}
\usepackage{epstopdf}
\usepackage[normalem]{ulem}
\usepackage{verbatim}
\usepackage{xcolor}
\definecolor{reviewgreen}{RGB}{0,128,0}

\usepackage{bm}

\def\red{\textcolor{red}}

\begin{document}
\title{Critical Non-Hermitian Skin Effect and Scale-Free Localization Morphing in Bilocally Coupled Hatano-Nelson Chains}

\author{Chong Wang}
\email{wangchong525@126.com}
\affiliation{Quantum Science Center of Guangdong-Hong Kong-Macao Greater Bay Area
(Guangdong), Shenzhen, China}
\affiliation{Guangdong Provincial Key Laboratory of Quantum Metrology and Sensing
\& School of Physics and Astronomy, Sun Yat-Sen University (Zhuhai
Campus), Zhuhai, China}

\author{Linhu Li}
\email{lilinhu@quantumsc.cn}
\affiliation{Quantum Science Center of Guangdong-Hong Kong-Macao Greater Bay Area
(Guangdong), Shenzhen, China}

\date{\today}
\begin{abstract}
We investigate the critical non-Hermitian skin effect (CNHSE) and the coupling-driven morphing of scale-free localization (SFL) in two Hatano–Nelson chains coupled locally at two bulk sites separated by an inclusive distance \(d\). At weak interchain coupling, competition between the opposite skin accumulations of the two chains generates a geometrically selected critical branch with a complex loop-like spectrum and scale-free skin states localized near the physical boundaries. Remarkably, this SFL branch persists as the coupling becomes strong, although its underlying mechanism and spatial profile change qualitatively. Strong local hybridization generates high-energy impurity states and effectively divides the low-energy Hilbert space into geometry-selected inner and outer segments, with the former supporting SFL states near the interchain-coupled sites and the latter retaining conventional skin states. The number and localization scale of the SFL states are controlled by the separation between the coupling sites. Consequently, increasing the interchain coupling continuously relocates the SFL states from the physical boundaries to the coupling-defined internal interfaces. Our results demonstrate that bilocal coupling can induce a CNHSE and further drive its scale-free states into a segmentation-induced localization regime.
\end{abstract}
\maketitle

\section{Introduction}

Recent advances in non-Hermitian physics have established dissipation and
nonreciprocity as useful resources for generating and controlling
unconventional physical phenomena, rather than merely detrimental effects to
be avoided~\cite{Bender1998RealSpectra,
Gong2018TopologicalPhases,Kawabata2019SymmetryTopology,lin2023topological,Ashida2020NonHermitianPhysics,
Bergholtz2021ExceptionalTopology}. A particularly prominent example is the
non-Hermitian skin effect (NHSE), in which an extensive number of eigenstates
become exponentially localized at system boundaries under open boundary
conditions (OBCs)~\cite{Hatano1996LocalizationTransitions,
Hatano1997VortexPinning,Lee2016AnomalousEdge,
MartinezAlvarez2018RobustEdge,Yao2018EdgeStates,
Kunst2018BiorthogonalBBC,Yokomizo2019NonBloch,
Lee2019AnatomySkinModes}. The resulting mismatch between Bloch spectra and OBC
eigenstates has motivated non-Bloch and generalized-Brillouin-zone
descriptions, while also revealing an extreme sensitivity to boundary
conditions and perturbations~\cite{Song2019ChiralDamping,
Okuma2020TopologicalOrigin,Zhang2020WindingSkin,Yang2020AuxiliaryGBZ,
Borgnia2020BoundaryModes,Longhi2019ProbingNHSE,Guo2021ExactGBC,PhysRevB.104.085401,
Wang2024Amoeba}.

The NHSE is not restricted to clean one-dimensional chains. Its interplay
with onsite dissipation, non-Bloch pumping, quasiperiodicity, domain walls,
higher-order topology, open-system dynamics, and many-body correlations
produces a broad family of spectral and dynamical phenomena~\cite{Yi2020OnSiteDissipation,
Longhi2020BandCollapse,Lee2020NonHermitianPumping,
Jiang2019AndersonLocalization,Deng2019DomainWall,
Lee2019HybridHigherOrder,Liu2020DynamicalCriticalSkin,
Ma2024ChiralNHSE,Jiang2024TunableNHSE,Hou2024DissolutionNHSE,
Peng2025LongRangeNHSE,Kim2024CollectiveNHSE,Yoshida2024MottNHSE}.
These ideas have also been explored experimentally in nonreciprocal mechanical
and acoustic structures~\cite{Brandenbourger2019RoboticMetamaterials,
Ghatak2020MechanicalMetamaterial,Li2024DynamicNHSE,
Xiong2024TrackingNHSE}, photonic and quantum simulators~\cite{
Xiao2020QuantumDynamics,Weidemann2020TopologicalFunneling,
Sun2024FloquetSkin,Liu2024ChiralEdgeSkin,Shen2025FermiSkin}, and
topolectrical circuits~\cite{Helbig2020TopolectricalBBC,
Hofmann2020ReciprocalSkin,Zhang2021HigherOrderNHSE,
Zou2021HybridHigherOrder}. Related circuit and ultracold-atom experiments
further demonstrate defect engineering, tunable nonreciprocal transport,
dynamical signatures of skin localization, and scale-free localization~\cite{
Stegmaier2021DefectEngineering,Gou2020NonreciprocalTransport,
Liang2022DynamicSignatures,Xie2024ObservationSFL,
Zhou2025AnomalousCircuitNHSE,Zhao2025TwoDimensionalNHSE}.

One particularly intriguing manifestation is the critical non-Hermitian skin
effect (CNHSE), characterized by strongly size-dependent spectral structures
and scale-free localization (SFL) of eigenstates~\cite{Li2020CriticalNHSE,
Yokomizo2021ScalingCNHSE,Qin2023CompetitiveScaling,Li2024UniversalScaleFree,
Cai2024DrivenCNHSE,Yi2025BoundaryDefectCNHSE,
Zhang2025MagneticCNHSE}. In its standard setting, the CNHSE emerges when two
subsystems with distinct skin localization properties are globally
coupled~\cite{Li2020CriticalNHSE}. Owing to the
exponential sensitivity associated with the NHSE, an arbitrarily weak
intersystem coupling can become nonperturbative in the thermodynamic limit,
reorganizing the spectrum and producing eigenstates whose localization
lengths scale with the system size~\cite{Li2020CriticalNHSE,
Yokomizo2021ScalingCNHSE,Qin2023CompetitiveScaling}. Closely related studies
have established impurity-induced SFL and its extensions to local
non-Hermiticity and disorder~\cite{Li2021ImpuritySFL,
Guo2023ScaleFreeAccumulation,Li2023ScaleFreePT,
Molignini2023AnomalousImpurity,Xie2024ObservationSFL,
Yilmaz2024SFLAnderson,Peng2025LongRangeNHSE}. The same exponential spectral sensitivity has
also motivated proposals for non-Hermitian sensing~\cite{
Budich2020TopologicalSensors,McDonald2020EnhancedSensing}.

Existing studies of the CNHSE have predominantly considered couplings that
extend throughout the system~\cite{Li2020CriticalNHSE,
Qin2023CompetitiveScaling}. Meanwhile, long-range asymmetric links have been
shown to select both the number and localization scale of a subset of states,
demonstrating that coupling geometry can itself become a controlling length
scale~\cite{Guo2024ScaleTailored}. This raises a natural question: can a finite
number of local couplings alone generate a critical skin response, and, if
so, how does the spatial geometry of these couplings determine the resulting
spectrum and eigenstates? Unlike global coupling, a finite number of local
links would conventionally be expected to act only as local perturbations in
an extended system. Their interplay with the intrinsic boundary sensitivity
of non-Hermitian skin states therefore provides a conceptually distinct
setting for exploring critical non-Hermitian localization.

Here, we address this question using a minimal model consisting of two one-dimensional Hatano-Nelson chains coupled at two spatially separated sites. We find that such bilocal coupling generates a geometrically selected critical branch characterized by complex, loop-like eigenenergy spectra and SFL states, whereas the remaining eigenstates retain real eigenenergies and conventional skin localization. Remarkably, the number of scale-free states is controlled by the separation between the two coupling positions, revealing a direct connection between the geometry of the local couplings and the structure of the critical spectral branch.

Even more intriguingly, increasing the bilocal coupling strength does not destroy the SFL branch, although its physical origin changes from critical hybridization to coupling-induced segmentation. Instead, the scale-free states undergo a pronounced localization morphing: their localization centers shift from the physical boundaries of the system to the two coupling positions, while their characteristic loop-like complex spectrum and scale-free nature persist. Apart from a small number of strongly localized defect states whose energies scale with the coupling strength, the remaining spectrum retains the same qualitative partition 
between the SFL loop branch and the NHSE real-line branch in regimes where they are well resolved. 
Meanwhile, a mixture between the two types of localization occurs in the regime where they possess analogous localization characteristics.
These results demonstrate that a finite number of local couplings can not only generate critical non-Hermitian behavior, but also provide a means of geometrically controlling and spatially relocating the eigenstates.

The remainder of this paper is organized as follows. In Sec.~II, we
introduce the bilocally coupled Hatano--Nelson model and use its
chain-exchange--reflection symmetry to reduce the full Hamiltonian to two
independent parity sectors. In Sec.~III, we characterize the NHSE real-line branch and the SFL loop branch through their spectra, eigenstate
profiles, and finite-size scaling, and examine effective segmentation at
strong coupling. Section~IV analyzes the mixing between NHSE and
SFL states at weak coupling and estimates the characteristic length scale
for the real-to-complex spectral transition. Section~V investigates the
dynamical signature of localization morphing under a slow ramp of the
bilocal coupling. Section~VI summarizes our results and their physical
implications.

\begin{figure}
\includegraphics[width=\columnwidth]{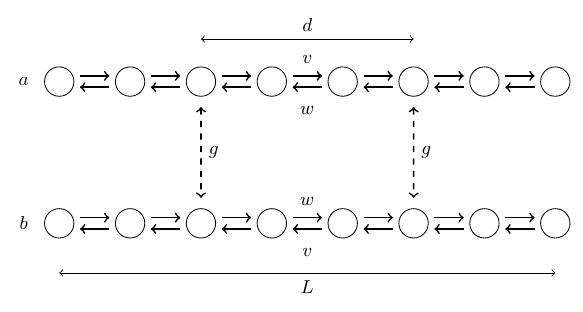}
\caption{Schematic of two locally coupled Hatano-Nelson chains. The two
chains, labeled by $a$ and $b$, have non-reciprocal nearest-neighbor
hopping amplitudes $v$ and $w$. They are coupled by a Hermitian inter-chain
coupling $g$ at two bulk sites separated by a distance $d$ in a chain of
length $L$.}
\label{fig:model-schematic}

\end{figure}

\section{Model}

We consider two bilocally coupled Hatano--Nelson chains~\cite{
Hatano1996LocalizationTransitions,Hatano1997VortexPinning} under OBCs,
as schematically illustrated in Fig.~\ref{fig:model-schematic}. 
The two chains are labeled by $s\in\{a,b\}$, and the tight-binding Hamiltonian can be written as
\begin{align}
	\hat{H}&=\hat{H}_{a}+\hat{H}_{b}+\hat{H}_{ab},\\
	\hat{H}_{s}&=\sum_{x=1}^{L-1}
	\left(t_{s,\rightarrow}\hat{s}_{x+1}^{\dagger}\hat{s}_{x}
	+t_{s,\leftarrow}\hat{s}_{x}^{\dagger}\hat{s}_{x+1}\right),\\
	\hat{H}_{ab}&=g\sum_{j=1}^{2}
	\left(\,\hat{a}_{x_{j}}^{\dagger}\hat{b}_{x_{j}}
	+\hat{b}_{x_{j}}^{\dagger}\hat{a}_{x_{j}}\right),
\end{align}
with $\hat{s}_{x}$  denoting the annihilation operators on chain $s$ at site $x$,
$t_{s,\rightarrow}$ and $t_{s,\leftarrow}$ the rightward and leftward hopping amplitudes, respectively,
and $g$ the Hermitian coupling between the two chains.
$\hat{H}_{ab}$ hybridizes the two chains at $x_1$ and $x_2$, resulting in an extra length scale $d=x_2-x_1+1$ to the system, in addition to the system's length $L$. For simplicity, we choose even numbers of $L$ and $d$ with spatially symmetric bilocal coupling,
\begin{equation}
	x_{1}=L/2-d/2+1,\qquad x_{2}=L/2+d/2.
\end{equation}
We further impose equal but opposite nonreciprocity on the two chains,
\begin{equation}
	t_{a,\rightarrow}=t_{b,\leftarrow}\equiv v,\qquad
	t_{a,\leftarrow}=t_{b,\rightarrow}\equiv w.
	\label{eq:opposite-nonreciprocity}
\end{equation}
The model thus satisfies a parity symmetry
\begin{equation}
	[H,\mathcal{P}]=0, \qquad\mathcal{P}^{2}=I,
	\label{eq:parity-symmetry}
\end{equation}
with $\mathcal{P}$ the symmetry operator, defined as
\begin{equation}
	\mathcal{P}|x,a\rangle=|L+1-x,b\rangle,
	\qquad
	\mathcal{P}|x,b\rangle=|L+1-x,a\rangle.
	\label{eq:parity-operation}
\end{equation}
Thus, the system can be block-diagonalized into decoupled subspaces with positive and negative parities,
which halves the matrix dimensions involved in numerics and thereby significantly reduces numerical errors induced by non-normality.
Specifically, the eigenstates can therefore be classified by
$\mathcal{P}|\Psi_{\sigma}\rangle=\sigma|\Psi_{\sigma}\rangle$,
where $\sigma=\pm1$. Equivalently, their amplitudes on the two original
chains satisfy
\begin{equation}
	b_{x}=\sigma a_{L+1-x}.
	\label{eq:parity-wavefunction-relation}
\end{equation}
The label $\sigma$ denotes the joint chain-exchange--inversion parity;
it is not a local bonding or antibonding combination at the same site.

To make the reduction explicit, we introduce the parity basis
\begin{equation}
	\begin{aligned}
		|x,\sigma\rangle&=\frac{1}{\sqrt{2}}
		\left(|x,a\rangle+\sigma|L+1-x,b\rangle\right),
		\\[-2pt]
		&\hspace{22mm}x=1,\ldots,L .
	\end{aligned}
	\label{eq:parity-basis}
\end{equation}
In this basis the Hamiltonian is block diagonal,
\begin{equation}
	\begin{aligned}
		H&\simeq H_{+}\oplus H_{-},\\
		\det(E-H)&=\det(E-H_{+})\det(E-H_{-}).
	\end{aligned}
	\label{eq:branch-factorization}
\end{equation}
Both sectors share the open-chain Hatano--Nelson Hamiltonian
\begin{equation}
	H_{0}=\sum_{x=1}^{L-1}
	\left(v|x+1\rangle\langle x|
	+w|x\rangle\langle x+1|\right).
	\label{eq:branch-bulk-Hamiltonian}
\end{equation}
The two local inter-chain couplings combine into a single reciprocal
bond between $x_{1}$ and $x_{2}$, with a parity-dependent sign. The
sector Hamiltonians are thus
\begin{equation}
	H_{\sigma}=H_{0}+\sigma g
	\left(|x_{1}\rangle\langle x_{2}|
	+|x_{2}\rangle\langle x_{1}|\right),
	\qquad \sigma=\pm1.
	\label{eq:branch-Hamiltonian}
\end{equation}
Equation~(\ref{eq:branch-Hamiltonian}) gives an exact reduction for any
finite $L$, any centered separation $d$, and arbitrary $g$. Each
sector contains $L$ eigenvalues, and their union gives the complete
$2L$-level spectrum.

The decomposition also enables a systematic analysis of the complex spectral features, as further discussed in Appendix~\ref{app:analytical-solution}.

\section{morphing of SFL from bilocal coupling}
\subsection{Skin modes and scale-free modes}
\begin{figure}
	\includegraphics[width=\columnwidth]{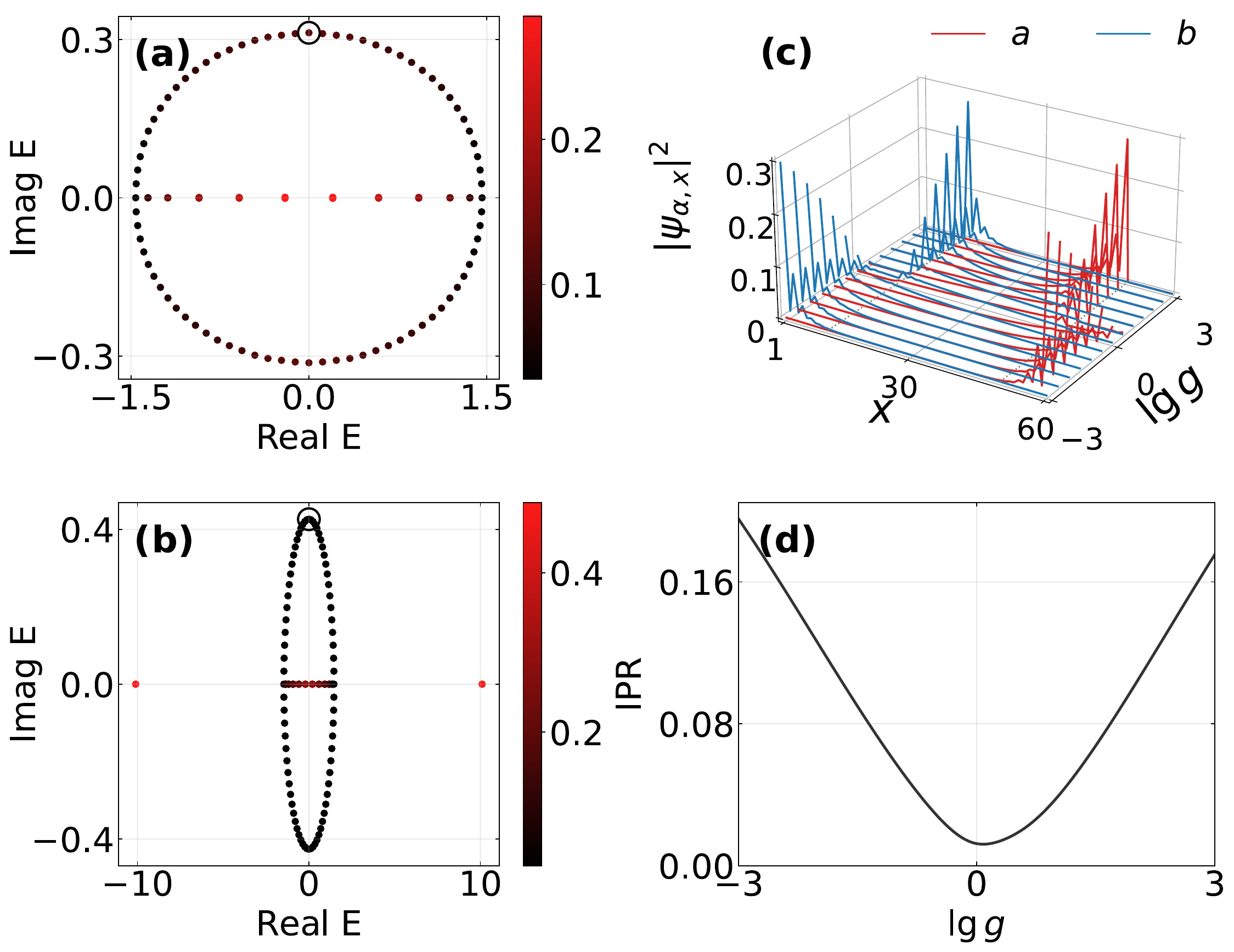}
	\caption{Spectral and eigenstate signatures of scale-free localization in the weak- and strong-coupling regimes. (a) and (b) Energy spectra at weak coupling
		$g=0.01$ and strong coupling $g=10$, respectively, colored by the IPR.
		Black circles mark the right eigenstate with the largest imaginary
		part of the eigenenergy in each spectrum.
		(c) Chain-resolved probabilities $|\psi_{\alpha,x}|^2$ of the right
		eigenstate selected independently at each $g$ by the same maximum-${\rm Im}E$
		criterion, plotted against the lattice site $x$ and $\lg g$ for
		$\lg g=-3,-2.5,\ldots,3$. Red and blue solid curves denote chains
		$a$ and $b$, respectively. The probability is normalized over both
		chains together. The profiles show the relocation of the selected
		state from the physical boundaries to the vicinity of the coupling sites.
		(d) IPR of the state selected independently at each $g$ by the same
		criterion.
		The parameters are $L=60$, $d=40$, $x_{1}=11$, $x_{2}=50$,
		$v=1.0$ and $w=0.5$.}
	\label{fig:cnhse-weak-strong}
\end{figure}

To begin with, we demonstrate a general picture of the localization morphing in Fig. \ref{fig:cnhse-weak-strong} by the spectral features and eigenstate profiles of two representative examples of the model, corresponding to the weak and strong bilocal-coupling regimes, respectively.
The energy spectra in Figs.~\ref{fig:cnhse-weak-strong}(a) and \ref{fig:cnhse-weak-strong}(b) are marked by the inverse participation ratio (IPR) of each eigenstate $|\psi_n\rangle=\sum_{\alpha=a,b}\sum_{x=1}^{L}
\psi_{\alpha,x}^{(n)}|x,\alpha\rangle$, defined as
\begin{equation}
	I_n=\sum_{\alpha=a,b}\sum_{x=1}^{L}
	\left|\psi_{\alpha,x}^{(n)}\right|^4
	=\sum_{\alpha=a,b}\sum_{x=1}^{L}
	\left(\rho_{\alpha,x}^{(n)}\right)^2,
\end{equation}
with $n$ labeling different eigenstates. Here $\rho_{\alpha,x}^{(n)}=|\psi_{\alpha,x}^{(n)}|^2$ is the site probability of a right eigenstate normalized by $\sum_{\alpha,x}\rho_{\alpha,x}^{(n)}=1$.
Due to the non-reciprocal hopping amplitudes, all eigenstates are localized, as expected for the systems with NHSE.
For the chosen parameters, the nondefect spectrum consists of an
NHSE branch near the real axis and an SFL branch forming a loop around it.
We refer to these as the {\it real-line} and {\it complex-loop} branches, respectively.
The latter describes the shape of the entire branch and does
not exclude the few real eigenenergies constituting the loop.
The eigenstates of the complex-loop branch have relatively smaller $I_n$; combined
with the complex eigenenergies, this resembles the characteristic signatures
of SFL~\cite{Li2020CriticalNHSE,Yokomizo2021ScalingCNHSE}, rather than skin localization in
decoupled Hatano-Nelson chains under OBCs.
Notably, 
these states are found to be localized
at the boundaries when $g \ll 1$, and at
the coupling positions when $g \gg 1$
[Fig. \ref{fig:cnhse-weak-strong}(c)],
indicating a localization morphing tuned by the bilocal couplings.
During the morphing process, the IPR in
Fig.~\ref{fig:cnhse-weak-strong}(d) first decreases and then
increases with $g$, meaning that the selected state evolves from localized
to spatially extended and then becomes localized again.

To reveal the different localization properties of the real-line and complex-loop branches, we analyze the normalized distribution averaged over the relevant eigenstate manifold, defined as
\begin{equation}
	\bar{\rho}_{\alpha,x}=
	\frac{\sum_{n\in\mathcal{E}}\left|\psi_{\alpha,x}^{(n)}\right|^2}
	{\max_{x,\alpha}\sum_{n\in\mathcal{E}}\left|\psi_{\alpha,x}^{(n)}\right|^2},
	\qquad \alpha\in\{a,b\} ,
	\label{eq:chain-resolved-normalized-average-density}
\end{equation}
where $\mathcal{E}$ denotes the eigenstate set included in the average (i.e., each of the two branches).

\begin{figure}[!t]
	\includegraphics[width=\columnwidth]{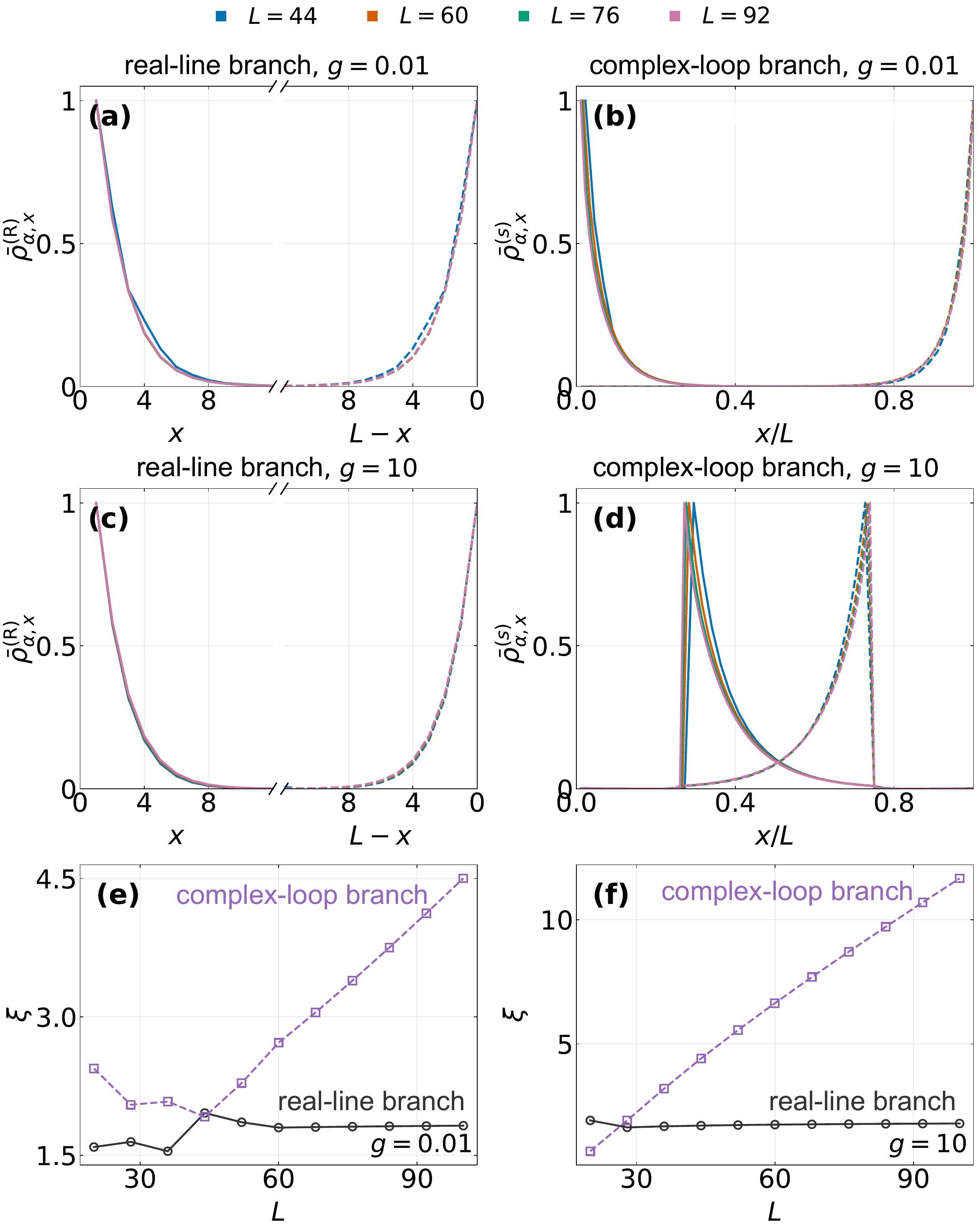}
	\caption{branch-resolved localization of the real-line and complex-loop
		states. Panels (a) and (b) show the real-line and complex-loop normalized
		average densities at $g=0.01$, and panels (c) and (d) show the corresponding
		profiles at $g=10$. Blue, orange, green, and magenta denote
		$L=44,60,76,92$, while dashed and solid curves denote chains $a$ and
		$b$, respectively. In the real-line panels, the broken horizontal axis
		shows the chain-$b$ profile against $x$ and the chain-$a$ profile against
		$L-x$ near their respective accumulating boundaries. Panels (e) and (f) show
		the fitted localization lengths at $g=0.01$ and $g=10$; black circles and
		purple squares denote the real-line and complex-loop branches, respectively. The fits
		use $L=20,28,\ldots,100$. Other parameters are $d=L/2$, $v=1.0$, and
		$w=0.5$.}
	\label{fig:scale-free-localization}
\end{figure}

At weak coupling, Figs.~\ref{fig:scale-free-localization}(a) and ~\ref{fig:scale-free-localization}(b) show that eigenmodes in
both branches retain opposite boundary accumulation on the two
chains, but with different scaling properties.
In this finite-size comparison, we vary $L$ while keeping $d=L/2$. Specifically, with spatial coordinate normalized to $x/L$, the complex-loop branch is seen to possess similar distribution profiles for different system sizes $L$, indicating a boundary SFL with localization length proportional to $L$~\cite{Li2020CriticalNHSE,Yokomizo2021ScalingCNHSE}.
In contrast, eigenmodes in the real-line branch show identical profiles for different $L$ in the original spatial coordinate, meaning that the localization length is fixed across different system sizes~\cite{Yao2018EdgeStates,Yokomizo2019NonBloch}.
The same localization features for the two branches are also seen in Figs.~\ref{fig:scale-free-localization}(c) and ~\ref{fig:scale-free-localization}(d), except that the complex branch has eigenstates distributed in the interval bounded by the two coupling sites, representing a defect SFL~\cite{Li2021ImpuritySFL,Xie2024ObservationSFL}.

To quantify this distinction, we extract the decay length of each
branch-resolved profile from
\begin{equation}
	\bar{\rho}_{\alpha,x}\sim
	A_{\alpha}\exp\left[-\frac{|x-x_{0,\alpha}|}{\xi_{\alpha}}\right],
	\label{eq:localization-length-fit}
\end{equation}
where the localization center $x_{0,\alpha}$ is chosen at the maximum
of the corresponding density profile. Depending on the coupling
strength and spectral branch, $x_{0,\alpha}$ can follow either the physical
boundary or the inner edge of the central segment near a coupling
site. The fitting procedure therefore follows the coupling-induced
relocation of the states. At weak coupling, both energy branches are
fitted over the full chain. At strong coupling, the real-line profile
is still fitted over the full chain, whereas the complex-loop fit is
restricted to the interval between $x_1$ and $x_2$. We also remove
states whose total probability on the four coupling-site basis states
exceeds $0.5$; this excludes the four strongly localized defect modes
at $g=10$. Moreover, the exchange--reflection symmetry maps the two
chain-resolved profiles onto each other and hence requires
$\xi_a=\xi_b$, so a single common localization length is sufficient
for each energy branch.

Figs.~\ref{fig:scale-free-localization}(e) and
\ref{fig:scale-free-localization}(f) reveal distinct size dependences
for the two branches. In the weak-coupling case, Fig.~\ref{fig:scale-free-localization}(e)
shows that the
two fitted localization lengths display nonmonotonic behavior and
become unstable around their crossing at small $L$. The physical origin
of this instability is analyzed in the next section.
Beyond this crossover, the real-line localization length approaches an
$L$-independent value, as expected for an ordinary skin mode controlled
by the microscopic hopping asymmetry~\cite{Yao2018EdgeStates,
Yokomizo2019NonBloch}, whereas the complex-loop
localization length grows approximately linearly with $L$. For
sufficiently large systems, Fig.~\ref{fig:scale-free-localization}(f) shows the same separation even
more clearly at strong coupling: the real-line localization length
remains nearly constant, while the complex-loop localization length
grows with the central-segment size $d=L/2$. This growth survives the
relocation of the localization centers away from the physical
boundaries. The simultaneous growth of $\xi$ and relocation of $x_0$
therefore show that increasing $g$ morphs a boundary SFL mode into a
coupling-induced defect SFL mode without destroying its scale-free
character. 

\subsection{Effective Segmentation in the Strong-Coupling Regime}
The persistence of scale-free localization at large $g$ can be
understood through effective segmentation. When the local inter-chain
coupling greatly exceeds the intra-chain hopping amplitudes, the
coupled sites $x_{1}$ and $x_{2}$ cease to behave as ordinary bulk
sites of two weakly hybridized Hatano--Nelson chains. At each coupling
position, $|x_j,a\rangle$ and $|x_j,b\rangle$ instead form strongly
hybridized bonding and antibonding combinations with leading energies
$\pm g$. These high-energy modes act as effective impurities, connecting
the strong-coupling limit to impurity-induced SFL mechanisms~\cite{
Li2021ImpuritySFL,Xie2024ObservationSFL}. States
whose energies remain on the scale set by $v$ and $w$ have little
weight on the hybridized sites, which consequently impose local
constraints on the low-energy dynamics.

This observation gives a real-space interpretation of the
strong-coupling regime. The two coupling sites partition the original
system into spatial segments that communicate only through virtual
transitions involving the high-energy bonding and antibonding modes.
The effective hopping across either coupled site is therefore
suppressed by $g$. In the limit
$|g|\gg |v|,|w|$, these sites act approximately as cuts, and the
low-energy wave functions reorganize into fragmented subsystems. The
selected loop-spectrum states are predominantly supported in the
central segment, while other low-energy states occupy the outer
segments and the defect modes remain concentrated near the coupling
sites. This interpretation is consistent with the profile in
Fig.~\ref{fig:cnhse-weak-strong}(c) and the central-segment distribution in
Fig.~\ref{fig:scale-free-localization}(d).

\begin{figure}
	\includegraphics[width=\columnwidth]{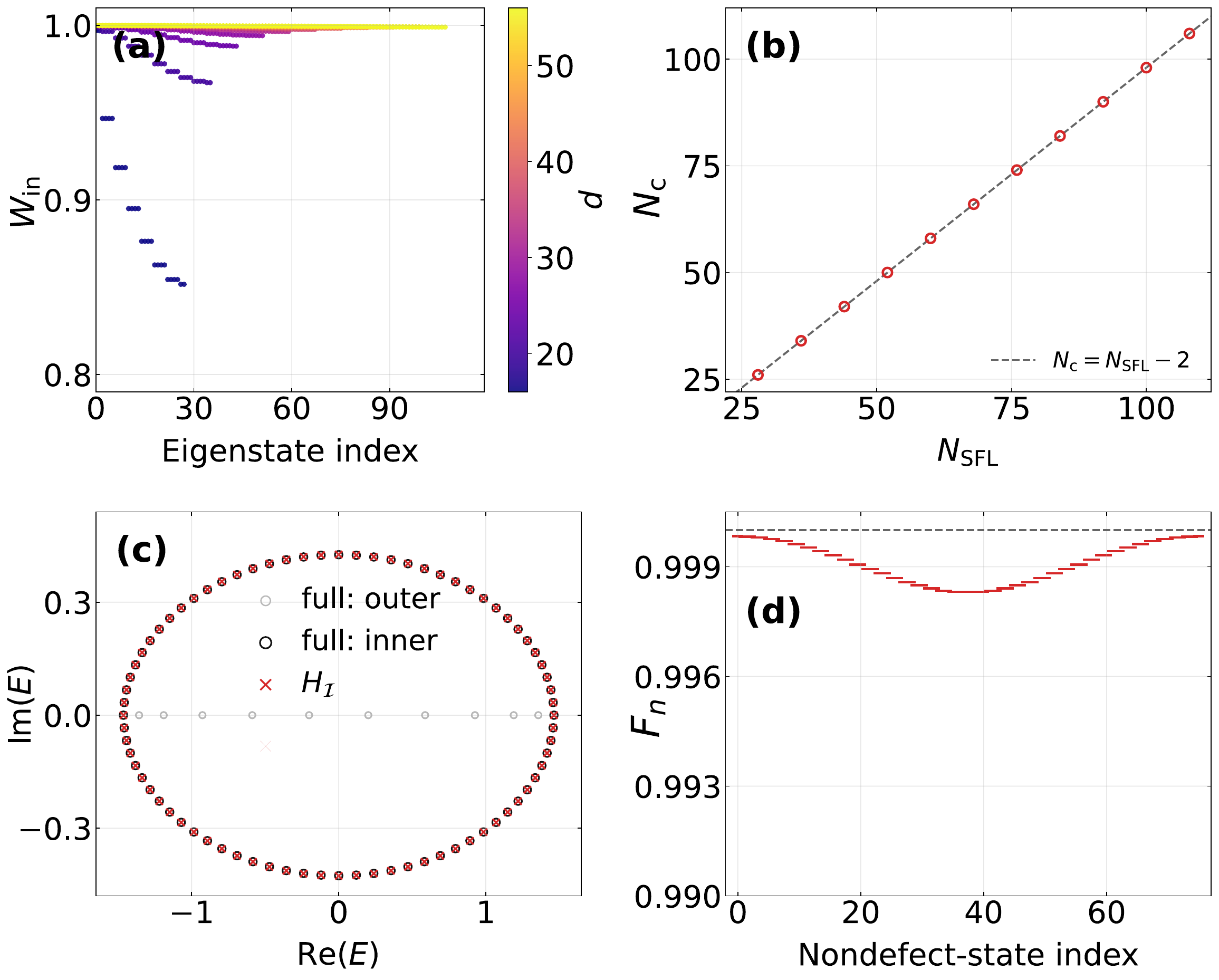}
	\caption{Counting and eigensystem evidence for effective segmentation in
		the strong-coupling regime. (a) Inner-segment weights
		$W_{\mathrm{in}}^{(n)}$ of the nondefect states selected by the
		inner-weight criterion; their number is $N_{\mathrm{SFL}}$.
		Different colors distinguish different values of $d$.
		(b) Spectral count $N_{\mathrm c}$ versus the
		position-selected count $N_{\mathrm{SFL}}$; the dashed line denotes
		$N_{\mathrm c}=N_{\mathrm{SFL}}-2$. (c) Complex spectrum of the full
		system compared with that of the isolated inner Hamiltonian
		$H_{\mathcal{I}}$ for $d=40$. Gray circles denote the
		outer states, black circles denote the position-selected inner-segment states,
		and red crosses are the eigenvalues of $H_{\mathcal{I}}$. The four
		high-energy defect eigenvalues lie outside the displayed window.
        (d) Single-state overlap $F_n$ for $d=40$ between
        full-system and inner-subsystem eigenstates matched by eigenenergy.
        The parameters are $L=60$,
		$d=16,20,\ldots,56$, $g=10$,
		$v=1.0$, and $w=0.5$.}
	\label{fig:effective-segmentation}
\end{figure}

The segmentation picture can first be tested directly by state
counting. We define the inner spatial segment
\begin{equation}
	\mathcal{I}
	=\left\{(\alpha,x)\,\middle|\,x_1\leq x\leq x_2,
	\ \alpha=a,b\right\},
\end{equation}
which contains the inclusive interval between the coupling points on
both chains. Its Hilbert-space dimension is $2d$. For the $n$th
normalized right eigenstate, the weight in this interval is
\begin{equation}
	W_{\mathrm{in}}^{(n)}
	=\sum_{\alpha=a,b}
	\sum_{x=x_1}^{x_2}
	\left|\psi_{\alpha,x}^{(n)}\right|^2 .
\end{equation}
We assign a state to the inner fragmented segment when
$W_{\mathrm{in}}^{(n)}>0.8$. This criterion selects $2d$ states over the
displayed range. Four are the high-energy defect states localized at the
coupling sites. 
After removing them, the remaining position-selected states
identify the SFL branch introduced above, with a total number given by
$$N_{\mathrm{SFL}}=2d-4$$
through an independent spatial criterion in the displayed regime.
{For numerical convenience, we also count all eigenstates whose imaginary parts exceed a threshold,}
\begin{align}
N_{\mathrm c}(d,L;g)=\#\{n:|\operatorname{Im}E_n|>\epsilon\},\qquad \epsilon=10^{-2}.
\label{eq:complex-mode-count-definition}
\end{align}
We note that $N_{\rm c}=N_{\rm SFL}-2$ is generally expected
as the complex loop coincides with the real axis in two points.

Fig.~\ref{fig:effective-segmentation}(a) shows the inner weights of the position-selected SFL states
after the four defect states are removed, and it is seen that a larger $d$ gives more states with $W_{\rm in}>0.8$.
Fig.~\ref{fig:effective-segmentation}(b) compares their number
$N_{\mathrm{SFL}}$ with the independently obtained $N_{\mathrm c}$.
Throughout the displayed range $d\geq16$, we find
exactly $N_{\mathrm c}=N_{\mathrm{SFL}}-2$ as expected.

The same inner segment provides a more direct eigensystem test. Let
$P_{\mathcal I}$ project onto $\mathcal I$. The isolated inner
Hamiltonian is the principal submatrix
\begin{equation}
	H_{\mathcal I}=P_{\mathcal I} H P_{\mathcal I}.
\end{equation}
It retains the central portions of both chains, the two local
inter-chain couplings, and the associated high-energy defect modes,
while removing only the hopping bonds from the coupling sites into
the outer segments.

To compare the spectra without relying on an arbitrary eigenvalue
ordering, let $n=1,\ldots,2d$ label the full-system eigenpairs
$(E_n,|R_n\rangle)$ selected by
$W_{\mathrm{in}}^{(n)}>0.8$, and let $\mu=1,\ldots,2d$ independently
label the eigenpairs
of $H_{\mathcal I}$, with eigenvalues $\varepsilon_{\mu}$. The symbol
$\pi$ denotes a permutation, or one-to-one map, from the full-system
label $n$ to the inner-subsystem label $\mu$; thus $\pi(n)$ is the
index of the inner-subsystem eigenvalue assigned to $E_n$. We choose
the optimal permutation by minimizing the total complex-plane distance
\begin{equation}
	\sum_{n=1}^{2d}\left|E_n-\varepsilon_{\pi(n)}\right|\label{eq:distance}
\end{equation}
over all possible permutations $\pi$. Fig.~\ref{fig:effective-segmentation}(c)
shows that the spectrum of $H_{\mathcal I}$ nearly coincides with 
the spatially selected full-system spectrum,
whereas the unselected outer eigenvalues form the complementary set.
For the displayed case $d=40$, the mean matched distance of Eq.~\eqref{eq:distance} is
$2.75\times10^{-3}$ and the maximum distance is
$4.95\times10^{-2}$.

To compare individual eigenstates, we exclude the four defect states from
both the selected full-system spectrum and the inner-subsystem spectrum.
We match these
remaining states by minimizing the total eigenenergy distance as above,
now restricting the assignment to the nondefect sets. Let
$|\widetilde R^{\mathcal I}_{\pi(n)}\rangle$ denote the normalized matched
inner-subsystem right eigenstate, embedded in the full lattice by setting
its amplitudes outside $\mathcal I$ to zero. For normalized $|R_n\rangle$,
the single-state overlap is
\begin{equation}
 F_n=\left|\langle\widetilde R^{\mathcal I}_{\pi(n)}|R_n\rangle\right|^2.
\end{equation}
Figure~\ref{fig:effective-segmentation}(d) gives a minimum overlap
of $0.998$ and a mean of $0.999$ for these nondefect states.
Thus, the simultaneous agreement of the inner-subsystem state count, spatially selected
eigenvalues, and right eigenstates gives direct evidence
that the strongly coupled sites spatially separate the central and outer segments of the lattice. At finite $g$ this
decomposition remains approximate because virtual processes through
the strongly hybridized sites produce residual coupling between the inner and outer segments.

\begin{figure}[!t]
	\includegraphics[width=\columnwidth]{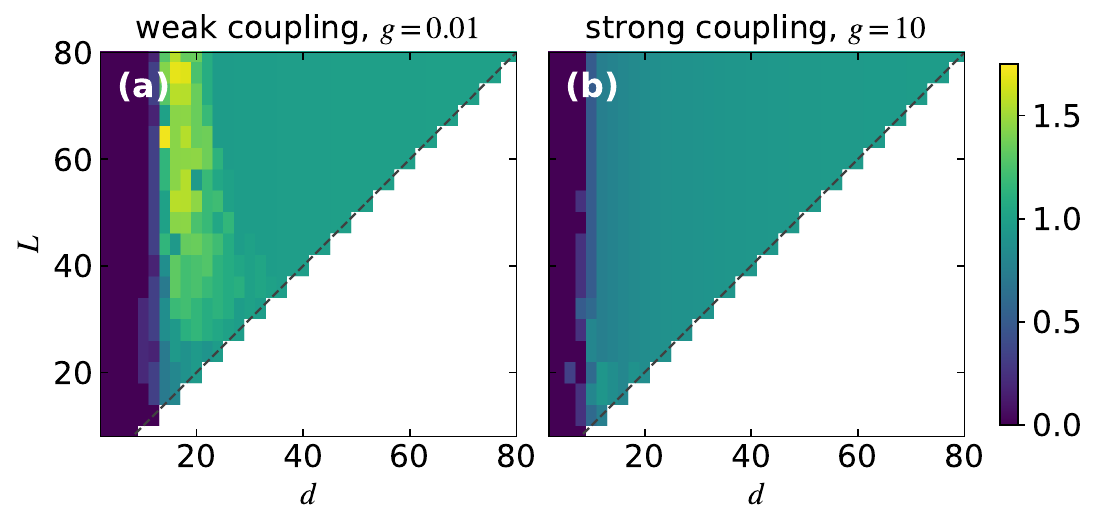}
	\caption{Complex-mode count in the $(d,L)$ plane at (a) weak coupling
		$g=0.01$ and (b) strong coupling $g=10$. The scan uses
		$L=8,12,\ldots,80$ and $d=2,4,\ldots,L$, with
		$v=1.0$ and $w=0.5$. The color gives
		$N_{\mathrm c}/(2d)$. The uncolored region $d>L$ is outside
		the centered two-coupling-point geometry, and the dashed line marks
		$d=L$. Both panels use the same color normalization.}
	\label{fig:complex-count-maps}
\end{figure}

These results distinguish the effective mechanisms underlying the
weak- and strong-coupling regimes.
In the weak-coupling regime, the two full Hatano-Nelson chains remain nearly
intact and the critical behavior originates from the competition
between their opposite skin accumulations. In the strong-coupling
regime, by contrast, the large local coupling first creates
impurity-like high-energy modes and then selects a reduced spatial
segment in which the remaining low-energy states develop scale-free
skin localization. 
On the other hand, we find that both mechanisms are governed by the coupling-point separation $d$, rather than the full-system size $L$, as shown by the count of $N_c$ versus different length scales in Fig. \ref{fig:complex-count-maps}.
In the strong-coupling regime, such a behavior can be attributed to the coupling-induced segmentation, where the inner segment can be viewed as a single PBC chain with $2d$ lattice sites and two strong impurities, providing a connection to impurity-induced SFL~\cite{Li2021ImpuritySFL,Guo2021ExactGBC}.
In the weak-coupling regime, the real-to-complex transition, and the count of complex eigenenergies, exhibit qualitatively similar trends regarding the variation with $d$ and $L$, suggesting that the CNHSE is also determined by the maximal distance between (weak) interchain couplings, rather than the full system size.

\section{Mixture between NHSE and SFL states in the weak-coupling regime}
\begin{figure}[!t]
	\includegraphics[width=\columnwidth]{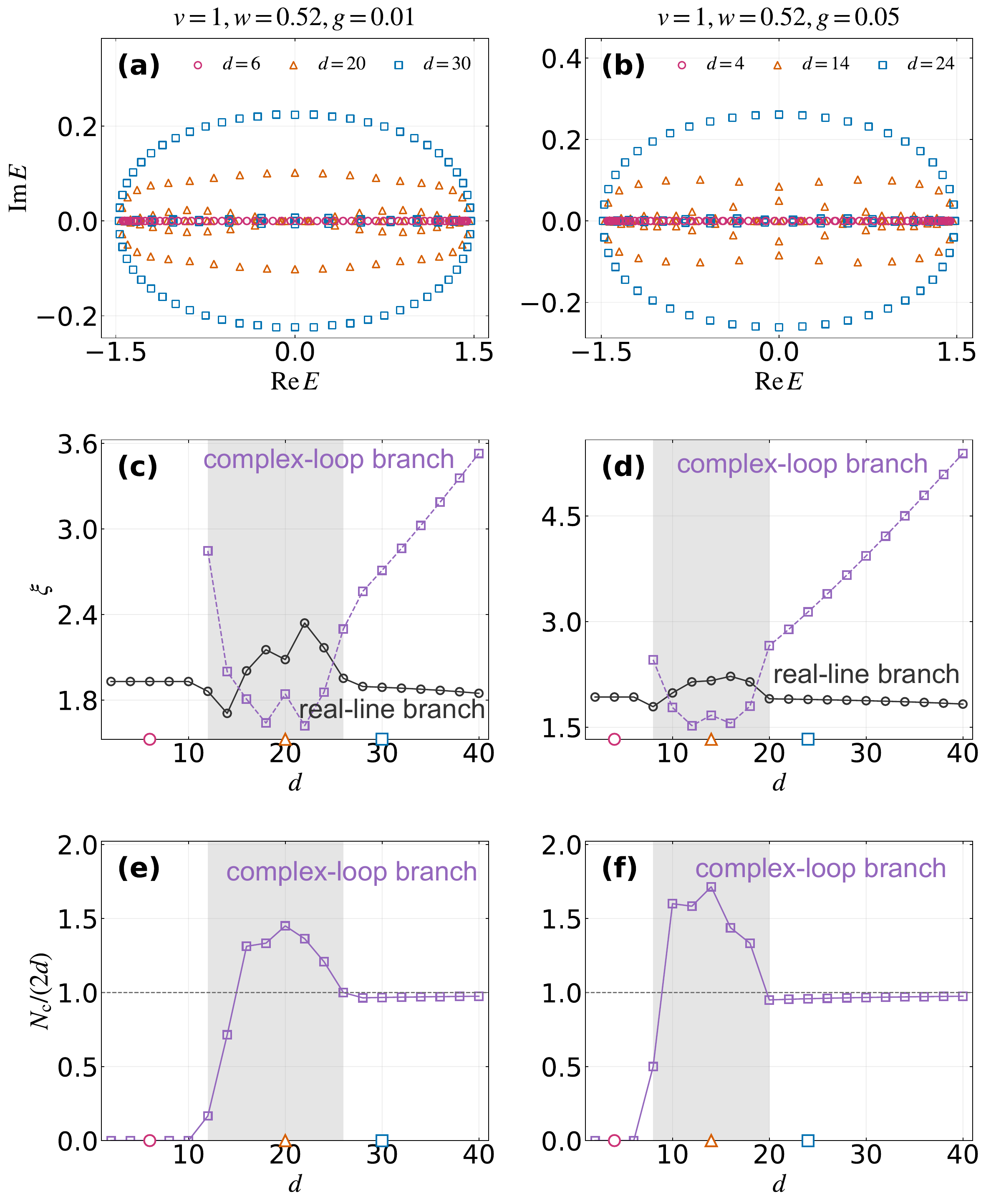}
	\caption{Spectral and localization signatures of the mixture between the
		NHSE and SFL branches in the weak-coupling regime. Panels (a) and (b) show the
		spectra at $L=60$, $v=1$, and $w=0.52$, with $g=0.01$ in the left
		column and $g=0.05$ in the right column. Purple-red circles, orange
		triangles, and blue squares denote $d=6,20,30$ in (a) and $d=4,14,24$
		in (b), respectively. Panels (c) and (d) show the fitted localization
		lengths of the real and complex branches versus $d=2,4,\ldots,40$.
		Panels (e) and (f) show the corresponding $N_{\mathrm c}/(2d)$ curves;
		the horizontal dashed line marks unity. Colored symbols on the horizontal
		axes identify the separations used in (a) and (b).
        The light-gray regions
		mark the intermediate NHSE--SFL mixing regime, spanning $12\le d\le26$
		in (c) and (e) and $8\le d\le20$ in (d) and (f). The unshaded regions
		to the left and right represent the effectively decoupled NHSE regime
		and the well-resolved CNHSE regime, respectively.}
	\label{fig:weak-coupling-mixture}
\end{figure}

In the previous discussion, we observe two anomalous behaviors in the weak-coupling regime. In Fig. \ref{fig:scale-free-localization}(e), the localization lengths of the real and complex branches are seen to be mixed for small $L$ (with $d=L/2$); in Fig. \ref{fig:complex-count-maps}(a), $N_c>2d$ is seen when $d$ is relatively small, meaning that the number of complex eigenenergies exceeds the size of the sub-Hilbert space of the central segment.
In addition, we further show the full spectrum with different $d$ in Figs. \ref{fig:weak-coupling-mixture}(a) and \ref{fig:weak-coupling-mixture}(b). It is seen that at an intermediate value of $d$, complex eigenenergies emerge also from the central real-line branch (orange triangles), but disappear at a larger $d$ where only the complex-loop branch supports nonzero Im$E$ (blue squares).
Such behaviors suggest a mixture between NHSE and SFL states, which become qualitatively indistinguishable in their spectral and localization features.

To further characterize these behaviors, in
Figs.~\ref{fig:weak-coupling-mixture}(c) to \ref{fig:weak-coupling-mixture}(f), we examine the localization and energetic properties with fixed system size $L=60$ and varying separation $d$ of the bilocal coupling,
at two different values of the coupling strength $g=0.01$ and $0.05$.
Three qualitatively different regimes can be identified accordingly.
\begin{itemize}
    \item An effectively decoupled NHSE regime: for a sufficiently small $d$, the two chains are effectively isolated from each other, possessing only real eigenenergies [red circles in Figs. \ref{fig:weak-coupling-mixture}(a) and \ref{fig:weak-coupling-mixture}(b)] and skin states with a size-independent localization length $\xi$.
    \item An intermediate NHSE-SFL mixing regime: at an intermediate separation $d$, complex eigenenergies emerge,  whose number varies with $d$ and generally satisfies $N_c>2d$. The localization lengths of real- and complex-energy states are mixed and do not show a clear trend with respect to $d$, suggesting a mixture between SFL ($\xi\sim L^1$) and NHSE ($\xi\sim L^0$).
    \item A well-resolved CNHSE regime: for a sufficiently large separation $d$ at the fixed weak coupling, the real-line and complex-loop branches are clearly distinguishable in both their spectral features and localization lengths, and $N_{\mathrm c}=N_{\rm SFL}-2=2d-2$ is recovered (note that $N_{\rm SFL}=2d$ in this regime due to the absence of defect states). 
\end{itemize}

The two boundaries between these regimes describe different changes. The first is a
real-to-complex spectral transition, marking the onset of complex
eigenvalues. The second is a transition
between the mixing regime and the well-resolved CNHSE regime, accompanied
by the recovery of $N_{\mathrm c}=2d-2$ and a clear distinction between
the two branches in the latter.

Finally, we note that as $d$ is reduced, the localization length of the SFL branch decreases and becomes comparable to the NHSE localization length [Figs.~\ref{fig:weak-coupling-mixture}(c) and ~\ref{fig:weak-coupling-mixture}(d)]. 
The two families then have strongly overlapping spatial
profiles and can no longer remain spectrally distinguishable, giving raise to the mixture between NHSE and SFL states. 
This interpretation is further verified by the absence of such a mixed regime in the strongly-coupled regime, where the NHSE and SFL are localized at spatially separated positions.
In such cases, we find that the real-line and complex-loop branches always remain clearly distinguishable (no shown), and the localization length for the latter remains linearly dependent on $d$ even when it drops below that for the former [Fig. \ref{fig:scale-free-localization}(f)].

\subsection{Characteristic length scales for the Real-to-complex spectraltransitions}
\label{sec:weak-coupling-estimate}
The real-to-complex spectral transition is a typical signature of a perturbative hybridization of the two chains.
To describe this transition,
we introduce a characteristic length scale $d_c$,
which characterizes the onset of complex
eigenvalues at weak coupling. This scale is controlled by the exponential enhancement
of a weak coupling between skin modes. A related
perturbative argument for boundary Floquet driving identifies an effective
hybridization scale proportional to
$|V|\exp[|\kappa_1-\kappa_2|L/2]$, where $V$ is the driving amplitude
and $\kappa_1$ and $\kappa_2$ are the signed inverse localization lengths
of the resonantly coupled modes~\cite{Hu2026BoundaryFloquet}.
There, a characteristic system size $L_c$ separates the real-spectrum
regime from the regime in which complex quasienergies emerge. Here the
coupling-point separation $d$, rather than the total system size, provides
the corresponding length scale.
In our static model, the corresponding exponential factor follows directly
from the parity-sector Hamiltonian in Eq.~(\ref{eq:branch-Hamiltonian}).
For $v>w>0$, we define the bulk hopping scale $J=\sqrt{vw}$ and
the nonreciprocity parameter
\begin{equation}
 \kappa=\frac12\ln\frac{v}{w},
 \label{eq:weak-coupling-skin-scale}
\end{equation}
where the amplitude decay length of an isolated-chain skin mode is
$\lambda_{\mathrm{amp}}=1/\kappa$. The decay length $\xi_\alpha$ in
Eq.~(\ref{eq:localization-length-fit}) is defined for the density rather
than the amplitude. For an exponential envelope, the density decay length
is half the amplitude decay length, $\xi_\alpha=\lambda_{\mathrm{amp}}/2$.
In the isolated-chain NHSE limit, this gives a localization length of NHSE
\begin{align}
\xi_{\mathrm{NHSE}}=1/(2\kappa).\label{eq:xi_NHSE}
\end{align}

The diagonal similarity transformation $S_{xx}=e^{\kappa x}$ makes
$S^{-1}H_0S$ an open reciprocal chain with hopping $J$, while the effective bond
between the coupling sites becomes
\begin{equation}
 \widetilde V_\sigma=\sigma g\left[
 e^{\kappa D}|x_1\rangle\langle x_2|
 +e^{-\kappa D}|x_2\rangle\langle x_1|\right],
 \quad D=d-1.
 \label{eq:gauged-effective-bond}
\end{equation}
Thus one direction of the effective bond is exponentially enhanced relative to
the reciprocal bulk hopping. This identifies a dimensionless scale
\begin{equation}
 u=\frac{|g|}{J}e^{\kappa(d-1)}.
 \label{eq:weak-coupling-control-parameter}
\end{equation}
The same exponential enhancement follows from the term linear in $g$
in the exact secular equation, as shown in
Eq.~(\ref{eq:secular-enhancement-asymptote}).

Estimating the onset of complex eigenvalues by $u\sim C$ gives the
characteristic length scale
\begin{equation}
 d_c\simeq1+\frac{1}{\kappa}\ln\frac{CJ}{|g|}
 \propto-\frac{\ln g}{\kappa}.
 \label{eq:weak-coupling-distance-estimate}
\end{equation}
Here $C$ represents the remaining spectral and geometric factors that must be determined from the full secular equation. At fixed $J$, weaker $g$ or a longer skin localization length shifts
this boundary to larger separation. Equation~(\ref{eq:weak-coupling-distance-estimate})
is a semiquantitative estimate of this boundary. 

\begin{figure}[!t]
 \includegraphics[width=\columnwidth]{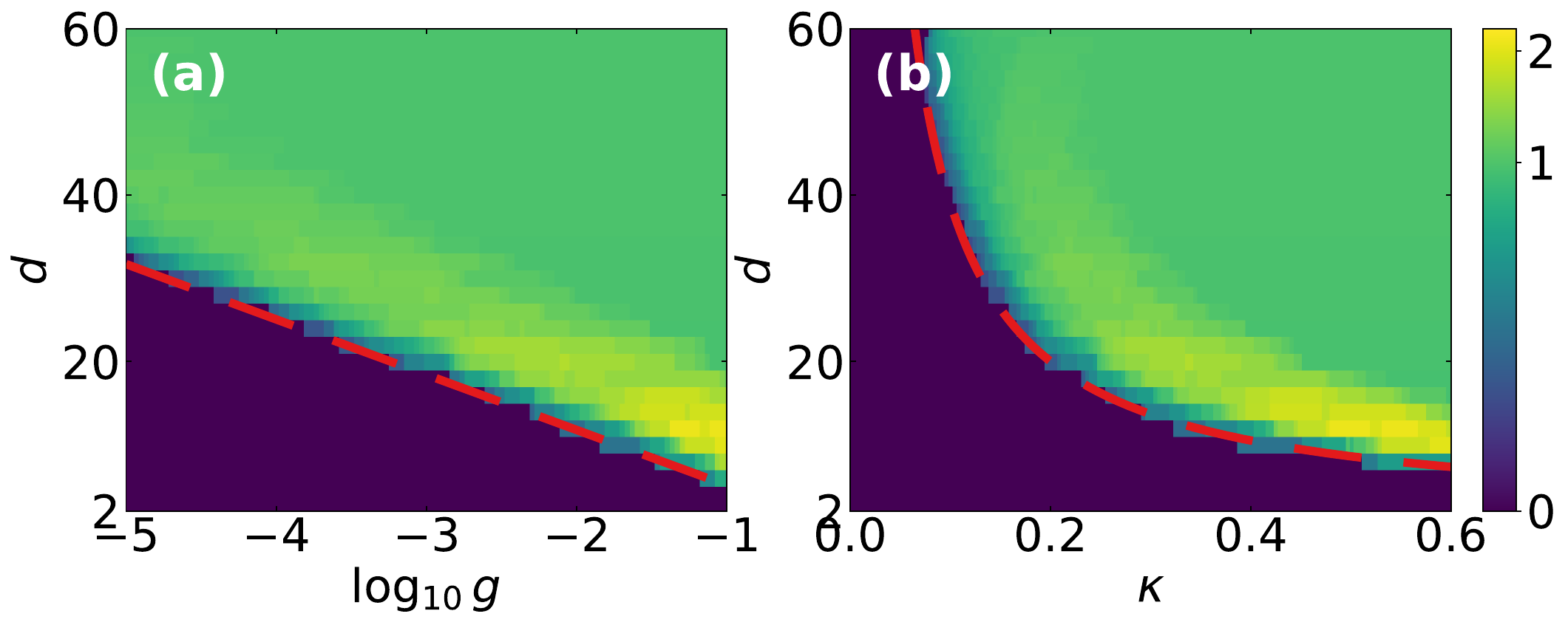}
 \caption{Onset of complex eigenvalues at weak coupling for $vw=1$ and $L=60$.
 (a) Complex-mode count $N_{\mathrm c}/(2d)$ in the
 $(\log_{10}g,d)$ plane at $\kappa=\tfrac12\ln2$.
 (b) The same quantity in the $(\kappa,d)$ plane at $g=0.01$.
 The hoppings are $v=e^\kappa$ and $w=e^{-\kappa}$, so both vary
 in (b) while the bulk energy scale $J=\sqrt{vw}=1$ remains fixed.
 Red dashed lines indicate the scaling
 estimate in Eq.~(\ref{eq:weak-coupling-distance-estimate}).
 They illustrate the logarithmic dependence
 on $g$ in (a) and the inverse dependence on $\kappa$ in (b).}
 \label{fig:weak-coupling-estimate}
\end{figure}

Figure~\ref{fig:weak-coupling-estimate} examines these trends while keeping
the bulk energy scale fixed by setting $vw=1$.
In Fig.~\ref{fig:weak-coupling-estimate}(a), the boundary between zero
and nonzero complex-mode counts moves toward smaller $d$ as $g$
increases. At fixed $g$, Fig.~\ref{fig:weak-coupling-estimate}(b) shows
the analogous shift as $\kappa$ increases and the intrinsic skin length
decreases. The red dashed lines follow the characteristic-length estimate
in Eq.~(\ref{eq:weak-coupling-distance-estimate}), with the multiplicative
factor $C$ adjusted for each parameter cut. They track the appearance of
complex eigenvalues and illustrate the logarithmic coupling dependence and
inverse-$\kappa$ distance scale in
Eq.~(\ref{eq:weak-coupling-distance-estimate}). 
  
The finite counting threshold in
Eq.~(\ref{eq:complex-mode-count-definition}) can shift the visible
boundary relative to the first emergence of a nonzero imaginary part.
The maps therefore support the predicted parameter dependence of the
characteristic length scale, which justifies that the real-to-complex transition originates from the same perturbative hybridization as for the CNHSE.

\section{Dynamical signature of the localization morphing}

As discussed above, in the weak-coupling regime, both the SFL and NHSE
states accumulate at the physical boundaries, whereas in the strong-coupling
regime, the two branches separate and the SFL states move into the central
segment bounded by the coupling sites. Varying the coupling strength can
therefore relocate a subset of the eigenstates. To track this relocation,
we compare the static eigensystem average with the time evolution
of a single initial eigenstate. Below we focus on chain $a$, since
chain $b$ is related to it by the exchange--reflection symmetry.

At an initial $g_i=10^{-2}$, we diagonalize the full Hamiltonian and select the right
eigenstate with the maximal imaginary energy as the initial state. We then
increase the coupling logarithmically according to
\begin{equation}
	g(t)=g_i\left(\frac{g_f}{g_i}\right)^{t/T},
	\qquad 0\leq t\leq T,
\end{equation}
with $g_f$ the final interchain-coupling strength at time $T$,
and solve
\begin{equation}
	\mathrm{i}\,\partial_t|\Psi(t)\rangle
	=H[g(t)]|\Psi(t)\rangle .
\end{equation}
Because non-Hermitian evolution does not conserve the Euclidean norm~\cite{
Ashida2020NonHermitianPhysics,Li2024DynamicNHSE}, the state is normalized
during propagation. This rescaling changes only
the overall amplitude and not the spatial profile. To use the same color
scale as the static map, we define the instantaneous normalized density
on chain $a$ as
\begin{equation}
	\rho^{\mathrm{dyn}}_{a,x}(t)
	=\frac{|\Psi_{a,x}(t)|^2}
	{\max_x|\Psi_{a,x}(t)|^2}.
	\label{eq:dynamical-chain-a-density}
\end{equation}

\begin{figure}[!t]
	\includegraphics[width=\columnwidth]{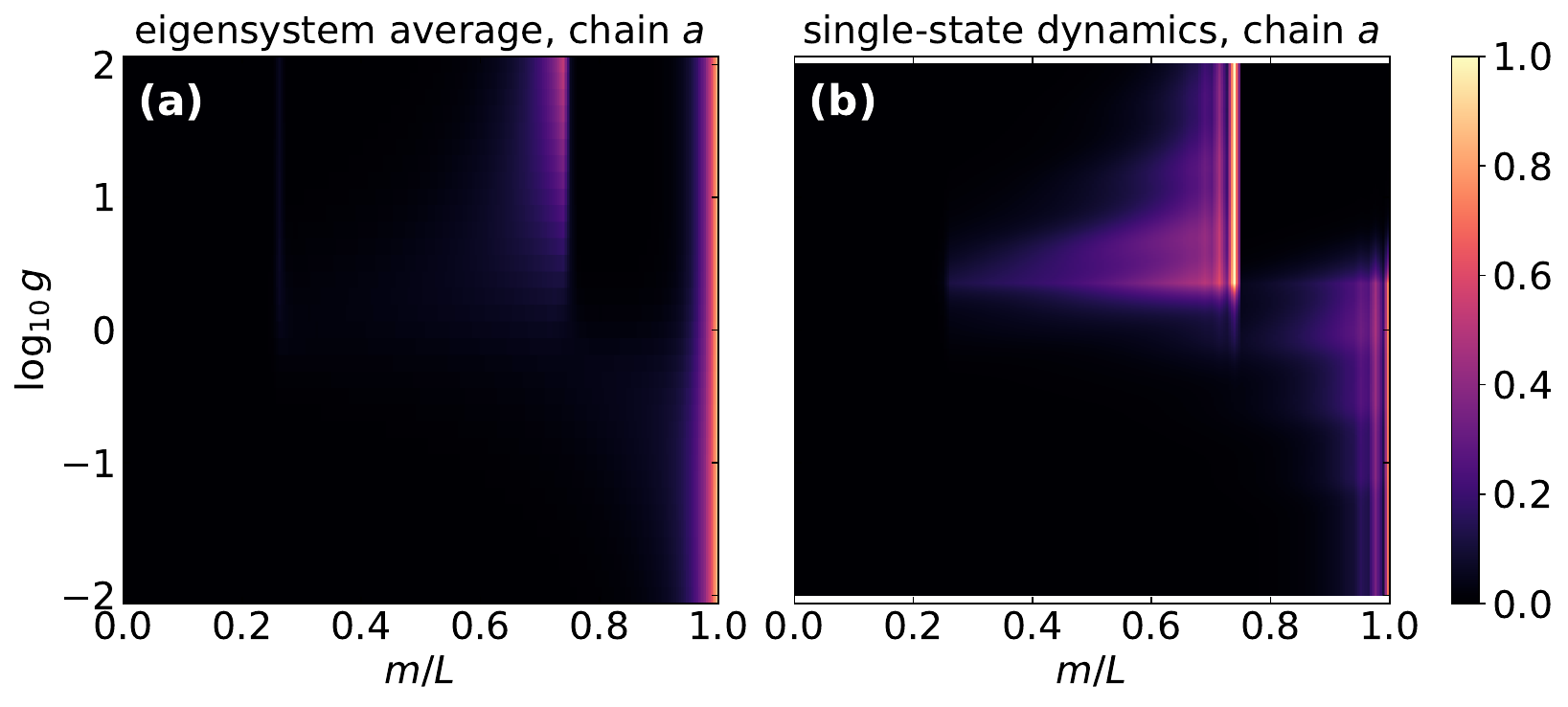}
	\caption{Static and dynamical coupling dependence on chain $a$.
		(a) Chain-$a$ normalized average density
		$\bar{\rho}_{a,x}$, obtained by independently diagonalizing the
		Hamiltonian and averaging over all right eigenstates at each $g$.
		(b) Chain-$a$ normalized density
		$\rho^{\mathrm{dyn}}_{a,x}$ during the slow evolution of a single
		complex edge eigenstate under a logarithmic ramp from $g_i=10^{-2}$
		to $g_f=10^2$ with $T=200$. In both panels $L=80$, $d=40$,
		$v=1.0$, and $w=0.5$. The full two-chain state is used
		in the propagation, while only its chain-$a$ component is shown.}
	\label{fig:complex-edge-dynamics}
\end{figure}

Fig.~\ref{fig:complex-edge-dynamics}(a) shows the coupling dependence of
the all-eigenstate average. As $g$ increases, the distribution separates
into two parts: one remains localized at the physical right boundary,
while the other moves inward toward the coupling position $x_2$ and enters
the central segment.

The single-state evolution in Fig.~\ref{fig:complex-edge-dynamics}(b)
provides the corresponding dynamical test. In the weak-coupling regime,
the chain-$a$ component remains localized near the physical right edge.
Across the intermediate-coupling regime, its dominant density moves
inward and reaches the right coupling position $x_2$. 
The evolution therefore gives a direct real-space manifestation of localization morphing and coupling-induced effective
segmentation: the complex edge state is transferred from the original
open boundary to an emergent boundary created by the local strong
coupling.

\section{Conclusion}

We have studied two Hatano--Nelson chains with opposite nonreciprocity
connected by Hermitian couplings at two spatially separated sites. This
minimal model supports a complex-loop SFL branch alongside a real-line NHSE
branch, showing that a finite number of local links can generate critical
non-Hermitian skin behavior.

In the weak-coupling regime, the two chains remain nearly intact. Their
opposite skin accumulations hybridize to form an SFL loop branch at the physical boundaries, including its two real-energy members in the resolved regime, while the separate real-line branch retains ordinary NHSE. The SFL
localization length scales with the system size, in contrast to the
 constant localization length of the NHSE states. When these two lengths
become comparable, finite-size hybridization blurs the separation between the
two spectral branches.

In the strong-coupling regime, the local bonds create four high-energy
bonding and antibonding defect modes and suppress low-energy propagation
through the coupled sites. The coupling sites then act as effective internal
boundaries that separate the central segment from the outer ones. The
complex-loop SFL states occupy the central segment, with their localization
scale and number set by the coupling-point separation $d$, whereas the
real-line NHSE states remain in the outer segments. The central-segment spectrum
and individual right eigenstates closely match their counterparts in the isolated inner Hamiltonian,
establishing the fragmentation mechanism.

 The SFL persists from weak to strong coupling despite the change in its underlying mechanism. As $g$ increases, the
SFL states move from the physical boundaries to the coupling-defined internal
interfaces. 
Bilocal coupling thus provides a geometry-controlled route to scale-free localization through distinct weak- and strong-coupling mechanisms.

\section*{ACKNOWLEDGMENTS}
L.L. acknowledges helpful discussion with Yu-Min Hu. This
work was supported by the National Natural Science Foundation
of China (Grant No. 12474159) and the Guangdong Provincial Quantum
Science Strategic Initiative (Grants No. GDZX2504003 and No. GDZX2504006).

\section*{DATA AVAILABILITY}
The data that support the findings of this article are not
publicly available upon publication because it is not technically
feasible and/or the cost of preparing, depositing, and
hosting the data would be prohibitive within the terms of this
research project. The data are available from the authors upon
reasonable request.

\appendix

\section{Analytical solution of the model}
\label{app:analytical-solution}

The parity reduction in Sec.~II turns the two-chain problem into two
independent chains, each with an additional reciprocal bond of strength
$\sigma g$. We use this representation to obtain the finite-size spectrum
and to identify the coupling enhancement discussed in the main text.

\subsection{Endpoint-coupled limit \texorpdfstring{$d=L$}{d=L}}

When $x_1=1$ and $x_2=L$, the additional bond joins the physical endpoints,
realizing a generalized boundary condition~\cite{Guo2021ExactGBC}.
Away from these endpoints, the right-eigenstate amplitudes obey
\begin{equation}
 E\psi_x=v\psi_{x-1}+w\psi_{x+1}.
 \label{eq:branch-bulk-recurrence}
\end{equation}
A component $\psi_x=z^x$ therefore has dispersion~\cite{
Yao2018EdgeStates,Yokomizo2019NonBloch}
\begin{equation}
 E=\frac{v}{z}+wz.
 \label{eq:branch-bloch-dispersion}
\end{equation}
For a given energy, the two roots $z_\pm$ give the bulk solution
\begin{equation}
 \psi_x=Az_+^x+Bz_-^x.
 \label{eq:branch-bulk-solution}
\end{equation}
The endpoint equations can be expressed in the same form as the bulk
recurrence by introducing auxiliary amplitudes $\psi_0$ and $\psi_{L+1}$:
\begin{equation}
 v\psi_0=\sigma g\psi_L,\qquad
 w\psi_{L+1}=\sigma g\psi_1.
 \label{eq:branch-generalized-boundary-condition}
\end{equation}
These auxiliary amplitudes do not represent additional physical sites.
Substituting the bulk solution gives the spectral condition
\begin{equation}
 \det\begin{pmatrix}
 v-\sigma g z_+^L & v-\sigma g z_-^L\\
 wz_+^{L+1}-\sigma g z_+ & wz_-^{L+1}-\sigma g z_-
 \end{pmatrix}=0.
 \label{eq:endpoint-branch-secular-equation}
\end{equation}
It selects energies for which the two endpoint conditions admit a
nonzero wave function. If $z_+=z_-$, the two independent bulk solutions
are $z^x$ and $xz^x$, rather than two identical powers.

\begin{figure}[H]
 \includegraphics[width=\columnwidth]{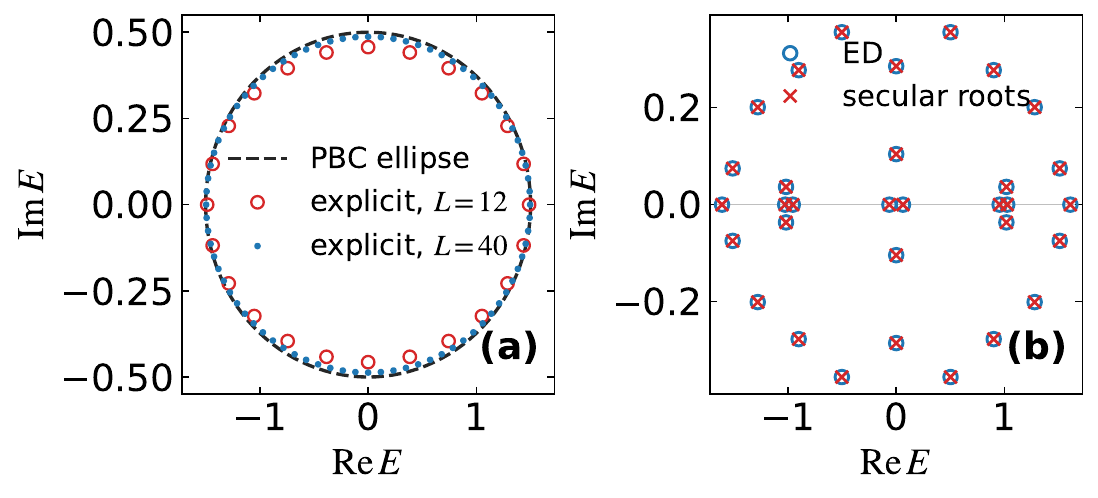}
 \caption{Explicit spectral limit and verification of the exact solution.
 (a) The finite-size spectra in Eq.~(\ref{eq:endpoint-explicit-spectrum})
 for $d=L$ and $g=J$, shown at $L=12$ and $40$, approach the PBC ellipse
 as $L$ increases. (b) Eigenenergies obtained from
 Eq.~(\ref{eq:explicit-branch-polynomial}) and from exact diagonalization
 (ED) of the original two-chain Hamiltonian for $L=16$, $d=10$, and
 $g=0.7$. Both cases use $v=1.0$ and $w=0.5$.}
 \label{fig:analytical-solution-checks}
\end{figure}

A simple explicit example is $g=J=\sqrt{vw}$, with $v>w>0$.
The endpoint conditions then admit a single component satisfying
\begin{equation}
 z^L=\sigma r,\qquad r=\sqrt{v/w}.
\end{equation}
Writing $\rho_L=r^{1/L}$ and
$k_{\sigma m}=[2\pi m+(1-\sigma)\pi/2]/L$, with $m=0,\ldots,L-1$,
gives the spectrum
\begin{equation}
 E_{\sigma m}=\frac{v}{\rho_L}e^{-ik_{\sigma m}}
             +w\rho_Le^{ik_{\sigma m}}.
 \label{eq:endpoint-explicit-spectrum}
\end{equation}
As $L$ increases, $\rho_L\to1$ and the allowed phases become dense.
The spectrum therefore approaches the PBC ellipse
$E_{\mathrm{PBC}}(k)=(v+w)\cos k+i(w-v)\sin k$~\cite{
Hatano1996LocalizationTransitions,Hatano1997VortexPinning}, as shown in
Fig.~\ref{fig:analytical-solution-checks}(a).

\subsection{Arbitrary coupling separation}

For $d<L$, the additional reciprocal bond connects two bulk sites.
Together with the original segment between them, it forms a loop with
two open tails. Ordinary sites still obey
Eq.~(\ref{eq:branch-bulk-recurrence}), while the coupling sites satisfy
\begin{align}
 E\psi_{x_1}&=v\psi_{x_1-1}+w\psi_{x_1+1}
                 +\sigma g\psi_{x_2},\nonumber\\
 E\psi_{x_2}&=v\psi_{x_2-1}+w\psi_{x_2+1}
                 +\sigma g\psi_{x_1}.
 \label{eq:bulk-coupling-matching}
\end{align}
We denote the tail lengths and the number of bonds between the coupling
sites by
\begin{equation}
 \ell_1=x_1-1,\qquad \ell_2=L-x_2,\qquad D=d-1.
 \label{eq:branch-length-definitions}
\end{equation}
Thus $d$ includes both coupling sites, whereas $D$ counts the intervening
bonds. For the centered geometry, $\ell_1=\ell_2=(L-d)/2$.

To express the spectrum compactly, let $p_n(E)$ be the characteristic
polynomial of an isolated open segment of $n$ sites. It satisfies
\begin{equation}
 p_0=1,\qquad p_1=E,\qquad
 p_n=Ep_{n-1}-vw\,p_{n-2}.
 \label{eq:open-chain-polynomial-recurrence}
\end{equation}
The added bond acts only on the two coupling sites. We introduce the
open-chain Green's function
\begin{equation}
 G_0(E)=(E-H_0)^{-1},\qquad
 \mathcal G_{ij}(E)=\langle x_i|G_0(E)|x_j\rangle.
 \label{eq:open-chain-Green-function}
\end{equation}
Only its values at the coupling sites are needed:
\begin{equation}
 \mathcal G(E)=\frac{1}{p_L}
 \begin{pmatrix}
 p_{\ell_1}p_{D+\ell_2}&w^Dp_{\ell_1}p_{\ell_2}\\
 v^Dp_{\ell_1}p_{\ell_2}&p_{D+\ell_1}p_{\ell_2}
 \end{pmatrix}.
 \label{eq:restricted-green-matrix}
\end{equation}
Here all polynomials are evaluated at $E$. The diagonal entries describe
return propagation to a coupling site, while the off-diagonal entries
describe propagation between the two sites through the open chain.
The reciprocal bond exchanges these sites, so its contribution gives
\begin{equation}
 D_\sigma(E)=p_L(E)\det[I_2-\sigma g\mathcal G(E)\tau_x],
 \qquad \tau_x=\begin{pmatrix}0&1\\1&0\end{pmatrix}.
 \label{eq:restricted-green-determinant}
\end{equation}
Evaluating this two-dimensional determinant gives the exact
characteristic polynomial of each parity sector,
\begin{equation}
 \begin{aligned}
 D_\sigma(E)={}&p_L(E)-p_{\ell_1}(E)p_{\ell_2}(E)\\
 &\times\left[\sigma g(v^D+w^D)+g^2p_{D-1}(E)\right].
 \end{aligned}
 \label{eq:explicit-branch-polynomial}
\end{equation}
Solving $D_\sigma(E)=0$ for $\sigma=\pm1$ gives all $2L$ eigenvalues.
The polynomial form also applies at the poles of $G_0$, where the
inverse in Eq.~(\ref{eq:open-chain-Green-function}) is not defined.
The factors involving $\ell_1$ and $\ell_2$ retain the influence of the
outer tails, so the bulk-coupled problem cannot be obtained by simply
replacing $L$ with $d$ in the endpoint result. For $d=L$, both tails
vanish and $p_{\ell_1}p_{\ell_2}=1$.

The term linear in $g$ also exposes the enhancement underlying the
weak-coupling analysis. Using $v=Je^\kappa$ and $w=Je^{-\kappa}$,
\begin{equation}
 v^D+w^D=2J^D\cosh(\kappa D).
 \label{eq:exact-coupling-enhancement}
\end{equation}
After measuring energies in units of $J$, the linear-coupling term
contains the dimensionless factor $2(g/J)\cosh(\kappa D)$. For $g>0$
and $\kappa D\gg1$,
\begin{equation}
 2\frac{g}{J}\cosh(\kappa D)
 \simeq\frac{g}{J}e^{\kappa D}=u.
 \label{eq:secular-enhancement-asymptote}
\end{equation}
This is the enhancement scale used in
Eq.~(\ref{eq:weak-coupling-control-parameter}). The remaining
energy-dependent polynomial factors determine the precise spectral
transition; the enhancement factor alone does not fix its location.

For each allowed energy, let
$|\phi_{\sigma n}^{R}\rangle=\sum_{x=1}^L\phi_{\sigma n,x}^{R}|x\rangle$
be a normalized right eigenstate of the reduced Hamiltonian $H_\sigma$.
The corresponding right eigenstate in the original two-chain basis is
\begin{equation}
 |\Psi_{\sigma n}^{R}\rangle
 =\frac{1}{\sqrt2}\sum_{x=1}^L\phi_{\sigma n,x}^{R}
 \left(|x,a\rangle+\sigma|L+1-x,b\rangle\right).
 \label{eq:reconstructed-right-eigenstate}
\end{equation}
Thus the second chain is obtained by reflection and the parity sign;
the characteristic polynomial determines the energies, while the
reduced eigenvalue equation determines the corresponding wave functions.

The roots of Eq.~(\ref{eq:explicit-branch-polynomial}) agree with direct
diagonalization of the full Hamiltonian in
Fig.~\ref{fig:analytical-solution-checks}(b), providing an independent
check of the numerical spectrum. The formulation also makes the
coupling geometry and the exponential enhancement explicit.

\bibliography{references}

\end{document}